\documentstyle[aps,prb,multicol,epsf]{revtex} 
\begin{document}
\draft
\title{Density dependence of the electronic supershells in the 
       homogeneous jellium model}
\author{Erik Koch\cite{byline} and Olle Gunnarsson}
\address{Max-Planck-Institut f\"ur Festk\"orperforschung, D-70569 Stuttgart}
\date{21 February 1996}
\maketitle 

\begin{abstract}
We present the results of self-consistent calculations of the electronic
shell and supershell structure for clusters having up to $6000$ valence
electrons. The ionic background is described in terms of a homogeneous
jellium. The calculations were performed for a series of different
electron densities, resembling Cs, Rb, K, Na, Li, Au, Cu, Tl, In, Ga, and Al,
respectively. By analyzing the occupation of the energy levels at the Fermi
energy as a function of cluster size, we show how the shell and supershell 
structure for a given density arises from the specific arrangement of energy 
levels. We investigate the electronic shells and supershells obtained for 
different electron densities. Using a scaling argument, we find a surprisingly 
simple dependence of the position of the supernodes on the electron density.
\end{abstract}
\pacs{71.24.+q, 36.40.Cg\hfill cond-mat/9606140}

\begin{multicols}{2}
\section{Introduction}

Following the observation of electronic shells\cite{Knight84} and the 
prediction of supershells,\cite{nishioka90} the electronic structure of 
metal clusters has attracted much interest.\cite{deHeerRev,BrackRev}
Treating the valence electrons as an ideal Fermi gas moving in a suitably 
chosen model potential, the observed structures can be well 
understood.\cite{Knight84,nishioka90,clemenger91} Such an effective 
one-electron model even lends itself to a semiclassical treatment. Using
a periodic orbit expansion\cite{BaBlo3,gutzwiller70} for the density of states,
the supershell structure can be interpreted as the beating pattern of the
contribution of two periodic orbits.\cite{nishioka90,bennemanntrafo,Lerme93a}
But obviously some arbitrariness is involved in choosing a ``suitable'' model
potential to describe a cluster. A systematic framework for determining 
such effective one-electron potentials is provided by density-functional
theory.\cite{HohenbergKohn,KohnSham,GunnarssonJones} Given the arrangement 
of the ions (Born-Oppenheimer approximation), the effective one-electron 
potential and the total energy of a cluster can be determined using the 
Kohn-Sham formalism. The microscopic structure of clusters is, however,
not known in general and the determination of the ground-state geometry 
from first principles is a major undertaking, which can only be performed for 
clusters made of not more than some ten atoms.\cite{Ballone,Rothlisberger,Jones}

Therefore drastic simplifications are called for to make self-consistent 
calculations for very large clusters feasible. In parallel to the early 
electronic 
structure calculations for solids\cite{Pines} and surfaces,\cite{LangKohnSurf}
one can, in a first approximation, describe the ionic background by a smooth
charge distribution, i.e.\ by a jellium. In the case of clusters the main
purpose of the jellium is to restrict the valence electrons to a finite volume,
that is to define the cluster size. Imagining a cluster to be a small drop it
seems reasonable to choose a spherically symmetric, homogeneous jellium. This 
leads to the homogeneous, spherical jellium model (HSJM).\cite{Beck,ekardt84}

It turns out that assuming sphericity of the cluster will essentially influence
only the amplitude of the shell oscillations. Spherical models tend to 
overestimate the amplitude of $\tilde{E}$ since there are no degrees of 
freedom allowing for static Jahn-Teller deformations in open-shell 
systems.\cite{Nilsson} Such effects give rise to a fine structure between the 
major shell minima. The latter, however, being related to spherical 
closed-shell clusters, are properly described by the model. Furthermore, it was 
recently shown that even clusters with a rough surface can be well described 
by a spherically averaged potential.\cite{Lerme95b} It thus appears that 
assuming spherical symmetry of the clusters for describing the electronic 
shell and supershell structure is justified. 

The homogeneity of the jellium, however, may well be questioned. It has turned
out that the shape of the jellium edge or, more generally, the shape of the
Kohn-Sham potential at the surface seems to influence the supershells strongly,
while leaving the electronic shells relatively unaffected.\cite{Lerme93b}
In particular for high-density metals large discrepancies between the 
predictions of the HSJM and the observed supershell structure have been 
observed, most notably in the case of gallium.\cite{Broyer93} 
Clearly the simple HSJM has to be refined, e.g.\ by choosing a suitable 
density profile at the surface and by including 
pseudopotentials.\cite{Lerme94,Lerme95a} In spite of that, the homogeneous
jellium model is still used as point of reference to compare to the results of
refined models. More importantly, a good understanding of this basic model
is needed to be able to find the relevant features that should be included in
an improved model, without introducing unnecessary complications.

The reason for including the noble metals Au and Cu is to bridge the
gap between the low density alkali metals and the high-density trivalent 
materials. It is clear that the presence of $d$ states will affect the
electronic shells. In the present work the results for these metals
mainly serve for uncovering the trends in the electronic supershells as
the electron density is varied.

The purpose of the present paper is threefold, which is reflected in its
organization. In Sec.~\ref{jellycalc}, we briefly review the homogeneous 
jellium model and present the results of our calculations. These include the 
oscillating part $\tilde{E}(N)$ of the total energy for densities resembling 
Cs, Rb, K, Na, Li, Au, Cu, Tl, In, Ga, and Al, respectively, for clusters
having 125 up to 6000 valence electrons. The position of the minima in 
$\tilde{E}$ (magic numbers) are listed explicitly. Furthermore, a method for 
determining the location of the nodes in the supershell oscillation is given
and the corresponding results for the clusters mentioned above are shown.
This collection of data could serve as reference for jellium results, 
extending previously published results considerably. 
\cite{Hintermann83,ekardt84,Beck,GenzkenPRL,Genzken93,Baguenard93,Lerme93b}

Focusing on the occupation of energy levels as the clusters increase in size,
a simple interpretation of the origin of supershell structure can be given.
This is done in Sec.~\ref{micro}. Analyzing the states at the Fermi 
energy, we show that the supershells arise from the fact that the
difference in angular momentum $\Delta l$ for orbitals of given number $n$ of
nodes in the radial wave function increases with cluster size. Since only
integer values of $\Delta l$ correspond to physical degeneracies, there is a 
periodic change from degenerate to nondegenerate energy levels as
the clusters grow larger. We give a semiclassical explanation of this 
observation, which provides an alternative to arguments using periodic 
orbit expansions.  

In Sec.~\ref{scaling} we analyze how the shell and supershell structure 
depends on the electron density. We show that for a homogeneous jellium edge
the softness of the potential at the surface is fairly independent of the 
density. It thus turns out that only the Wigner-Seitz radius $r_s$ 
(related to the electron density by $3/(4\pi r_s^3)$) is the relevant length 
scale for the electronic structure. We actually find that the supershells are 
linearly shifted with $1/r_s$, while the magic numbers are practically 
independent of the electron density. This suggests a general mechanism of 
{\em how} the surface softness influences the supershell structure. Finally, 
a fit to our data allows the results of jellium calculations to be 
estimated without actually having to perform the lengthy computations. 

Throughout the paper we give lengths in Bohr radii ($a_0$) and energies in 
Rydbergs ($Ry$).

\section{Jellium Calculations}\label{jellycalc}
\subsection{Method}

In the homogeneous jellium model a given material is characterized by the
average density $n_{bulk}$ of the valence electrons. Later on, it will be 
convenient to use the Wigner-Seitz radius $r_s$, in terms of which the bulk 
electron density is given by $n_{bulk}=3/(4\pi r_s^3)$. To make the description
of the clusters independent of the valence of the metal under consideration, we 
will use the number $N$ of valence electrons, {\em not} the number of atoms, to 
denote the size of a cluster. In the spherical, homogeneous jellium model
the jellium density is then given by $n_I(r)=n_{bulk}\,\Theta(R_0-r)$, where 
$R_0=r_s\,N^{1/3}$ is the radius of the cluster.

To find the total energy $E(N)$ of a jellium cluster having $N$ valence 
electrons we solve the Kohn-Sham equation
\begin{equation}\label{KSeqn}
  \Big(-\Delta + V_{KS}(\vec{r}\,)\Big)\Psi_\mu(\vec{r}\,) 
= \epsilon_\mu \Psi_\mu(\vec{r}\,)
\end{equation}
self-consistently. The Kohn-Sham potential is given by the sum of the
electrostatic potentials arising from the jellium and the electronic 
charge density, respectively, and the exchange-correlation potential
\begin{displaymath}
V_{KS}(\vec{r}\,)= V_I(\vec{r}\,) 
                 + 2\int\,d^3r'\,{n_{el}(\vec{r}\,')\over|\vec{r}-\vec{r}\,'|}
                 + V_{xc}(\vec{r}\,) .
\end{displaymath}
The electron density $n_{el}(\vec{r}\,)$ is determined from the $N$ 
lowest-lying eigenfunctions of (\ref{KSeqn})
\begin{displaymath} 
 n_{el}(\vec{r}\,) = \sum_{\mu:occ} |\Psi_\mu(\vec{r}\,)|^2 .
\end{displaymath}
Having reached self-consistency, the total energy is given by
\begin{eqnarray}
 E(N) = \sum_{\mu:occ}\epsilon_\mu &&
      - \int d^3r\,V_{KS}(\vec{r}\,)n_{el}(\vec{r}\,) \nonumber\\
      &&+ E_{xc}[n] + E_{Coul}[n] ,
\end{eqnarray}
where $E_{xc}$ is the exchange-correlation energy functional and $E_{Coul}$
is the electrostatic self-energy of the total (both jellium and electronic) 
charge density.

To treat large clusters more efficiently, instead of actually finding all the
eigenfunctions of the Kohn-Sham equation, we determine the electron density 
$n_{el}(r)$ and the sum $\sum\epsilon_\mu$ of the one-electron energies by
contour integration from the one-electron Green's function of (\ref{KSeqn}).
Exploiting spherical symmetry, the observables are given by
\begin{eqnarray}
 n_{el}(r)=&& \sum_l {\textstyle{2(2l+1)\over4\pi r^2}\,{1\over2\pi i}}
            \oint\limits_{E_F} g_l(E;r,r)\,dE \label{nel}\\[2ex]
 \sum_\mu\varepsilon_\mu=& 
         \displaystyle{4\pi\!\!\int\limits_0^\infty\!dr\,r^2\;}&
         \!\sum_l {\textstyle{2(2l+1)\over4\pi r^2}\,{1\over2\pi i}}
            \oint\limits_{E_F} g_l(E;r,r)\,E\,dE \label{SumE} .
\end{eqnarray}
Here $g_l(E;r,r)$ is the trace of the Green's function for the radial Kohn-Sham
equation with angular momentum $l$. It is constructed from the two solutions
$u_l^<(E;r)$ and $u_l^>(E;r)$ which are regular at the center and at infinity,
respectively
\begin{displaymath}
  g_l(E;r,r) = -{u_l^<(E;r)\;u_l^>(E;r) \over W(E)} ,
\end{displaymath}
where $W(E) = {u^<_l}'(E;r)\,u^>_L(E;r) - u^<_l(E;r)\,{u^>_l}'(E;r)$
is the Wronskian of (\ref{KSeqn}).

At zero temperature the integration contour would have to go through the
Fermi energy $E_F$ and we would have to deal with poles on the real axis.
We therefore introduce a fictitious finite but small temperature and multiply 
the Green's function in Eqs.~(\ref{nel}) and (\ref{SumE}) by a Fermi-Dirac
distribution. The contour can then be closed well above $E_F$, where the poles
on the real axis have negligible weight. The Fermi-Dirac distribution, however,
has additional poles at $E_F + (2n+1)\pi\,i\,kT$, for integer $n$. Their 
contributions have to be subtracted from the contour integral.
Working with fractional occupations has the 
further advantage of reducing oscillations in the occupation numbers 
during the iterative process of reaching self-consistency.

The main reason for using the Green's function approach to solving the
Kohn-Sham equation is its computational efficiency for large clusters.
Since the radial mesh grows with the cluster radius $R_0=r_s\,N^{1/3}$,
the computation of the Green's function scales with $N^{1/3}$. To determine
$n_{el}$ and $\sum\epsilon_\mu$ we have to perform $l_{max}\propto R_0$
integrations, for which the integration contour is independent of $N$. Hence 
the computer time grows as $N^{2/3}$. This has to be compared to the 
traditional approach, which for each $l$ involves finding $n_{max}\propto R_0$
eigenstates, thus resulting in a computational complexity of ${\cal O}(N)$.

Because of the the large density gradients at the cluster surface, it seems 
highly desirable to go beyond 
local-density approximation to the exchange-correlation energy. Fig.~\ref{DFT} 
shows the self-consistent Kohn-Sham potentials for a 
Na$_{5000}$ jellium obtained using local density approximation 
and two different generalized gradient approximations 
(GGA's).\cite{PW_GGA_test,EV_GGA}
Unfortunately, it turns out that the potentials found using a GGA exhibit
an unphysical shoulder at the surface. We therefore stick to the local-density
approximation\cite{VWN_LDA} for all subsequent calculations.

\subsection{Results}

We have performed self-consistent calculations for jellium clusters with 
electron densities $n_{bulk}$ corresponding to Cs, Rb, K, Na, Li, Au, Cu, Tl,
In, Ga, and Al (see Table \ref{rs_tbl}). Cluster sizes range from 125 to 6000 
valence electrons for each density considered. The calculations yield the total
energy $E(N)$ as a function of cluster size.

Neglecting the discreteness of energy levels (i.e., disregarding shell effects) 
the total energy is approximated by the asymptotic expansion
\begin{equation}\label{Esmooth}
 \bar{E}(N) = a_1\,N + a_2\,N^{2/3} + a_3\,N^{1/3} + \cdots.
\end{equation}
In this Thomas-Fermi-like expansion the leading coefficient 
$a_1$ resembles the bulk energy of the system. The next 
coefficient $a_2$ is related to the surface energy.\cite{PerdewClust} 
The electronic shells and supershells reflect the deviations
of the total energy $E(N)$ from this smooth function. These are given by
the oscillating part of the total energy, defined by
\begin{equation}\label{Eosc}
  \tilde{E}(N) = E(N) - \bar{E}(N).
\end{equation}

\noindent
\begin{minipage}{3.375in}
\begin{figure}
  \centerline{\epsfxsize=3.37in \epsffile{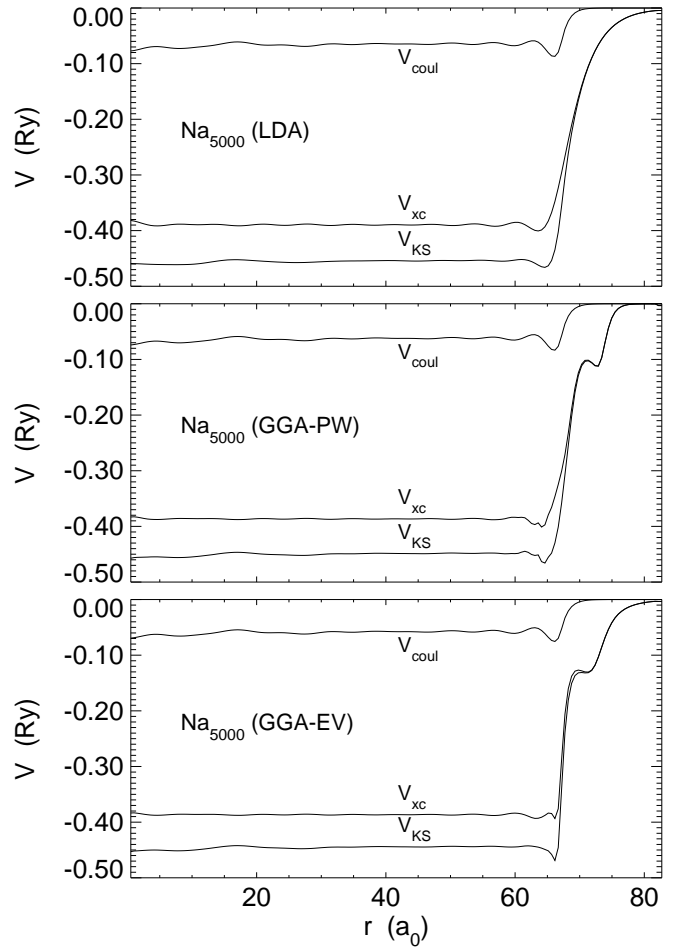}}
  \vspace{3ex}
  \caption[]{\label{DFT} 
            Self-consistent Kohn-Sham potentials for a homogeneous, spherical 
            Na$_{5000}$ jellium ($V_{KS}=V_{coul}+V_{xc}$). The results shown
            here differ only in the way the exchange-correlation term was 
            treated. The potential displayed in the first plot (LDA) was 
            obtained using the local-density approximation in the 
            parametrization given by Vosko, Wilk, and Nusair 
            (Ref.\cite{VWN_LDA}). The second (GGA-PW)
            and third (GGA-EV) plot show the potentials obtained using 
            generalized gradient approximations as given by 
            Perdew-Wang {\it et al.} (Ref.\cite{PW_GGA_test}) and Engel and
            Vosko (Ref.\cite{EV_GGA}), respectively.}
\end{figure}
\end{minipage}

\vspace{4ex}
Thus, to extract $\tilde{E}(N)$ from the total energy $E(N)$ obtained from 
our self-consistent calculations we need to know the smooth part $\bar{E}(N)$
of the total energy. To this end we have determined the parameters $a_i$ in 
the asymptotic expansion (\ref{Esmooth}) by a least-squares fit. To check the 
quality of our fit, we compare the fit parameters to known properties of 
jellium systems. The leading parameter $a_1$ should be equal to the bulk 
energy per electron of a homogeneous, neutral electron gas
\begin{displaymath}
 \epsilon_{bulk} = {3\over5}k_F^2 + \epsilon_{xc}(r_s) .
\end{displaymath}
The comparison of the parameters $a_1$ from the fit to $\epsilon_{bulk}$ is
shown in Fig.~\ref{ebulk}. 

\noindent
\begin{minipage}{3.375in}
\begin{figure}
  \centerline{\epsfxsize=3.37in \epsffile{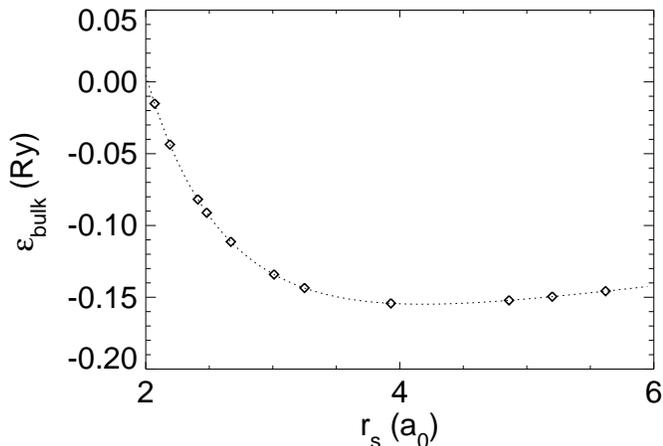}}
  \vspace{1ex}
  \caption[]{\label{ebulk}
             Leading term in the asymptotic expansion of the total energy
             $E(N)$ extracted from our calculations for homogeneous jellium
             spheres of densities $n_{el}=3/(4\pi r_s^3)$. For comparison the
             dotted line shows the energy of a homogeneous electron gas 
             in local density approximation using the parametrization
             given by Vosko, Wilk, and Nusair (Ref.\cite{VWN_LDA}).}
\end{figure}
\end{minipage}

\vspace{3ex} 
The next to leading term $a_2\,N^{2/3}$ should resemble the total energy 
of the jellium surface, hence $a_2$ should be related to the surface energy 
$\sigma$ by
\begin{equation}
  \sigma={a_2\over4\pi r_s^2} .
\end{equation} 
Fig.~\ref{esurf} shows the comparison between the surface energy obtained
from the fit parameter $a_2$ and the surface energies for jellium surfaces
calculated by Lang-Kohn\cite{LangKohnSurf} and 
Perdew {\it et al.}\cite{PW_GGA_test}.
The discrepancies between the surface energies found by Lang-Kohn and the other
data are probably due to the use of a different exchange-correlation energy
functional: Lang and Kohn used the Wigner approximation to the 
exchange-correlation energy, while the treatment of exchange and correlation
in the work of Perdew et al.\ and in our calculations was based on the results
of Monte Carlo simulations of the homogeneous electron gas. \cite{CA-egas}

\begin{figure*}
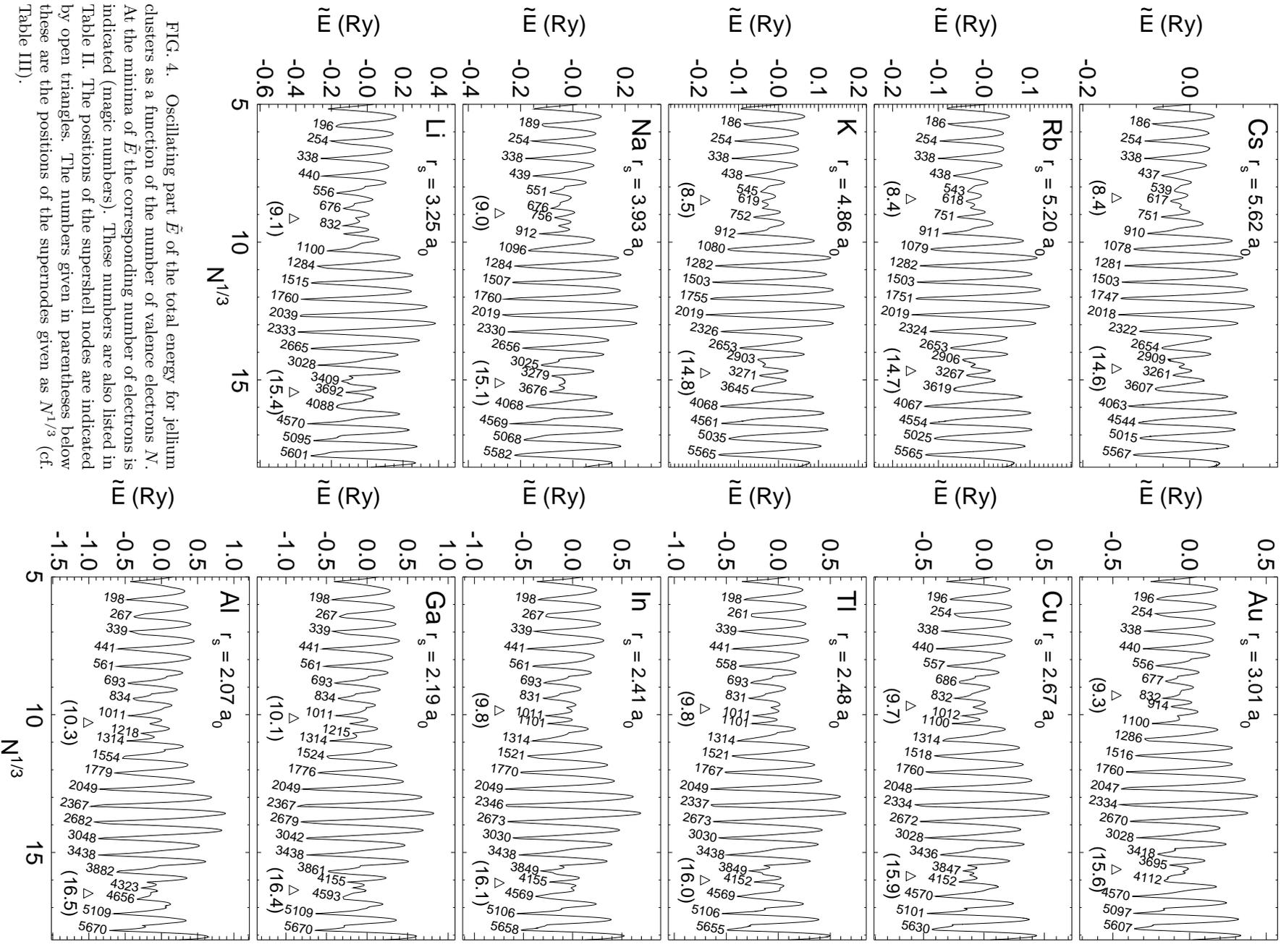

  \begin{minipage}{3.375in}
    \centerline{\epsfxsize=3.32in \epsffile{jellyosci1.epsi}}
    \vspace{3ex}
    \caption[]{\label{jellyosci}
             Oscillating part $\tilde{E}$ of the total energy for jellium
             clusters as a function of the number of valence electrons $N$.
             At the minima of $\tilde{E}$ the corresponding number of electrons
             is indicated (magic numbers). These numbers are also listed in
             Table \ref{magic_tbl}. The positions of the supershell
             nodes are indicated by open triangles. The numbers given in
             parentheses below these are the positions of the supernodes given 
             as $N^{1/3}$ (cf.\ Table \ref{snode_tbl}).}
    \vspace*{0.20cm}
  \end{minipage}
  \begin{minipage}{3.375in}
    \centerline{\epsfxsize=3.32in \epsffile{jellyosci2.epsi}}
  \end{minipage}
\end{figure*}

Having found the parameters of the asymptotic expansion (\ref{Esmooth}) we can
readily determine the oscillating part $\tilde{E}(N)$ of the total energy. 
The results from our self-consistent calculations for the densities listed in
Table \ref{rs_tbl} are shown in Fig.~\ref{jellyosci}. Plotting $\tilde{E}$
over $N^{1/3}$, which is proportional to the cluster radius, we find regular
oscillations (shell structure). The amplitude of these oscillations is 
modulated, resembling a beating pattern (supershell structure). 

The local minima in the oscillating part $\tilde{E}$ of the total energy 
correspond to clusters that are expected to be exceptionally stable. We call 
the number of valence electrons $N$ in these clusters {\em magic numbers}. They
are listed in Table \ref{magic_tbl}. 

\noindent
\begin{minipage}{3.375in}
\begin{figure}
  \centerline{\epsfxsize=3.37in \epsffile{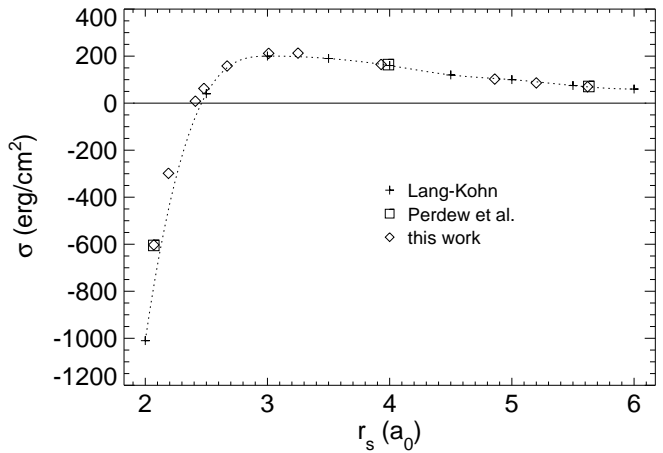}}
  \vspace{1ex}
  \caption[]{\label{esurf}
             Next to leading term in the asymptotic expansion of the total
             energy $E(N)$ extracted from our calculations for homogeneous
             jellium spheres of densities $n_{el}=3/(4\pi r_s^3)$ (diamonds).
             For comparison the surface energies $\sigma$ for jellium surfaces
             as found by Lang and Kohn (Ref.\cite{LangKohnSurf}, Table II) and 
             by Perdew {\it et\ al.} (Ref.\cite{PW_GGA_test}, Table XII) are 
             shown. The dotted line gives an interpolation of the Lang-Kohn 
             data.}
\end{figure}
\end{minipage}

\vspace{3ex}
It turns out that in the regions where the
amplitude of the shell oscillations is large, the magic numbers for different 
jellium densities are comparable. Comparing the sequences of corresponding 
minima in $\tilde{E}$, we find that the magic numbers grow slightly as the 
Wigner-Seitz radius decreases. In the regions where the amplitude of the shell 
oscillations is small, similar sequences of magic numbers exist. Here, however,
they do not extend over the whole range of jellium densities. Instead we find 
that a sequence starting at small densities ends at some intermediate density, 
while a new sequence, shifted by about half a period, evolves and extends
to larger jellium densities. 

For all jellium densities we find a pronounced modulation in the amplitude 
of the shell oscillations. Even a superficial look at Fig.~\ref{jellyosci}
shows that the beating pattern in $\tilde{E}$ is shifted towards larger
cluster sizes as the jellium density increases. To characterize the supershell
structure we use the location of the minima in the amplitude of the
shell oscillations. The position of these supernodes is given by minima
in the envelope of $\tilde{E}(N)$. The envelope is obtained eliminating
the rapid shell oscillations from 
$|\tilde{E}(N^{1/3})|$ using a low-pass Fourier filter. Naturally such a 
filtering scheme introduces an uncertainty in $N^{1/3}$ which is of the order 
of a period of the shell oscillation.

For all jellium densities we have considered, exactly two supernodes fall in
the range of clusters having 125 -- 6000 valence electrons. Using the procedure
described above, we have determined the position of these supernodes. The
results are listed in Table \ref{snode_tbl}.
\end{multicols}

\begin{table*}
\begin{tabular}{p{8ex}|ddddddddddd}
&{\large Cs} &{\large Rb} &{\large K}  &{\large Na} &{\large Li} &{\large Au} 
&{\large Cu} &{\large Tl} &{\large In} &{\large Ga} &{\large Al}\\[1ex]
\hline\\[-2ex]
 $r_s\;\;(a_0)$ &5.62&5.20&4.86&3.93&3.25&3.01&2.67&2.48&2.41&2.19&2.07
\end{tabular}
  \caption[]{\label{rs_tbl}
             Wigner-Seitz radii $r_s$ for the different metals which were used
             in the self-consistent calculations to define the jellium density.}
\end{table*}

\vspace{-2ex}
\begin{table*}
\begin{tabular}{p{8ex}rrrrrrrrrrr}
&{\large Cs} &{\large Rb} &{\large K} &{\large Na} &{\large Li} &{\large Au} &{\large Cu} &{\large Tl} &{\large In} &{\large Ga} &{\large Al}\\[1ex]\hline\\[-2ex]
&   137 &   137 &   138 &   138 &   138 &   138 &   138 &   138 &   138 &   138 &   138\\
&   186 &   186 &   186 &   189 &   196 &   196 &   196 &   198 &   198 &   198 &   198\\
&   254 &   254 &   254 &   254 &   254 &   254 &   254 &   261 &   267 &   267 &   267\\
&   338 &   338 &   338 &   338 &   338 &   338 &   338 &   339 &   339 &   339 &   339\\
&   437 &   438 &   438 &   439 &   440 &   440 &   440 &   441 &   441 &   441 &   441\\
&   539 &   543 &   545 &   551 &   556 &   556 &   557 &   558 &   561 &   561 &   561\\
&   617 &   618 &   619 &   634 &   640 &   638 &   638 &   639 &   639 &   639 &      \\
&   668 &   673 &   674 &   676 &   676 &   677 &   686 &   693 &   693 &   693 &   693\\
&   751 &   751 &   752 &   756 &   759 &   783 &       &       &       &       &      \\
&       &       &       &   832 &   832 &   832 &   832 &   831 &   831 &   834 &   834\\
&   910 &   911 &   912 &   912 &   912 &   914 &   921 &   924 &   927 &   927 &      \\
&       &       &       &       &       &  1013 &  1012 &  1011 &  1011 &  1011 &  1011\\
&  1078 &  1079 &  1080 &  1096 &  1100 &  1100 &  1100 &  1101 &  1101 &  1101 &  1101\\
&       &       &       &       &       &       &       &       &       &  1215 &  1218\\
&  1281 &  1282 &  1282 &  1284 &  1284 &  1286 &  1314 &  1314 &  1314 &  1314 &  1314\\
&  1503 &  1503 &  1503 &  1507 &  1515 &  1516 &  1518 &  1521 &  1521 &  1524 &  1554\\
&  1747 &  1751 &  1755 &  1760 &  1760 &  1760 &  1760 &  1767 &  1770 &  1776 &  1779\\
&  2018 &  2019 &  2019 &  2019 &  2039 &  2047 &  2048 &  2049 &  2049 &  2049 &  2049\\
&  2322 &  2324 &  2326 &  2330 &  2333 &  2334 &  2334 &  2337 &  2346 &  2367 &  2367\\
&  2654 &  2653 &  2653 &  2656 &  2665 &  2670 &  2672 &  2673 &  2673 &  2679 &  2682\\
&  2909 &  2906 &  2903 &  2913 &       &       &       &       &       &       &      \\
&  3070 &  3063 &  3056 &  3025 &  3028 &  3028 &  3028 &  3030 &  3030 &  3042 &  3048\\
&  3261 &  3267 &  3271 &  3279 &  3283 &  3288 &       &       &       &       &      \\
&       &       &       &       &  3409 &  3418 &  3436 &  3438 &  3438 &  3438 &  3438\\
&       &       &       &  3510 &       &       &       &       &       &       &      \\
&  3607 &  3619 &  3645 &  3676 &  3692 &  3695 &       &  3708 &  3708 &       &      \\
&       &       &       &       &       &  3865 &  3847 &  3849 &  3849 &  3861 &  3882\\
&  4063 &  4067 &  4068 &  4068 &  4088 &  4112 &  4152 &  4152 &  4155 &  4155 &  4158\\
&       &       &       &       &       &       &       &  4332 &  4320 &  4317 &  4323\\
&  4544 &  4554 &  4561 &  4569 &  4570 &  4570 &  4570 &  4569 &  4569 &  4593 &  4656\\
&  5015 &  5025 &  5035 &  5068 &  5095 &  5097 &  5101 &  5106 &  5106 &  5109 &  5109\\
&       &       &       &       &       &       &  5471 &       &       &  5511 &  5502\\
&  5567 &  5565 &  5565 &  5582 &  5601 &  5607 &  5630 &  5655 &  5658 &  5670 &  5670\\ 
\end{tabular}
  \caption[]{\label{magic_tbl}
             List of the magic numbers for jellium clusters of different
             densities. Listed are the numbers of valence electrons $N$ in
             the range from $125$ to $6000$ for which the oscillating part
             of the total energy has a minimum. The magic numbers for the
             different jellium clusters are arranged in sequences of
             comparable size. In the regions where the amplitude of the
             shell oscillations is large (cf.\ Fig.~\ref{jellyosci})
             these sequences extend over the whole range of densities.
             In the vicinity of the nodes in the supershell structure
             this is no longer true. Here sequences starting at small 
             jellium densities end at intermediate densities, while
             a new sequence, shifted by about half a period, evolves and
             extends to larger densities. 
            }
\end{table*}

\vspace{-2ex}
\begin{table*}
\begin{tabular}{p{8ex}|ddddddddddd}
&{\large Cs} &{\large Rb} &{\large K}  &{\large Na} &{\large Li} &{\large Au} 
&{\large Cu} &{\large Tl} &{\large In} &{\large Ga} &{\large Al}\\[1ex]
\hline\\[-2ex]
 $N^{1/3}_1$& 8.39& 8.43& 8.47& 8.95& 9.15& 9.30& 9.68& 9.79& 9.85&10.13&10.29\\
 $N^{1/3}_2$&14.59&14.67&14.75&15.11&15.45&15.62&15.86&16.01&16.11&16.37&16.49 
\end{tabular}
  \caption[]{\label{snode_tbl}
 Position of the nodes in the supershell structure for jellium clusters
 of different densities. The position of the supernodes was determined
 by finding the minima in the envelope of $\tilde{E}(N)$. $N$ denotes the
 number of valence electrons in the cluster.}
\end{table*}

\begin{multicols}{2}
\section{Microscopic Analysis}\label{micro}

A cluster with $N$ valence electrons will be more stable than neighboring 
clusters if its highest occupied orbital is completely filled. The increase
in stability can be characterized by the gap between the highest
occupied and the lowest unoccupied level. In general, the width of this gap 
rapidly decreases with growing cluster size. A set of degenerate levels will
however open a larger gap in the spectrum. The electronic shell structure
can thus be traced back to degeneracies in the energy spectrum of the cluster
potential at the Fermi energy $E_F$. In the following we therefore focus 
on the Kohn-Sham energies $\epsilon_{n,l}$ next to $E_F$. In doing so 
we identify the degeneracies that cause the strong shell 
oscillations and find a mechanism that explains why there are no such
degeneracies near the supernodes. 

The potentials in our self-consistent calculations are spherically symmetric. 
Therefore we can characterize the energy levels by two quantum numbers: 
the number $n$ of nodes in the radial wave function and the
angular momentum quantum number $l$. The degeneracy of an eigenenergy 
$\epsilon_{n,l}$ is given by $2(2l+1)$. Hence states with large $l$
have a stronger influence on the shell structure than those with small 
angular momentum. 

Since in our self-consistent calculations the Kohn-Sham levels are populated
according to a Fermi-Dirac distribution
\begin{displaymath}
  N_{\epsilon_{n,l}} = {2(2l+1)\over1+e^{(\epsilon_{n,l}-E_F)/kT}}
\end{displaymath}
degenerate levels $\epsilon_{n,l}$ near the Fermi energy $E_F$ are filled 
simultaneously. By ``filling'' a level we mean that the occupation of 
$\epsilon_{n,l}$ changes with cluster size (i.e.\ with the total number $N$
of valence electrons):
\begin{displaymath}
  \Delta N_{\epsilon_{n,l}}(N) 
       = N_{\epsilon_{n,l}}(N) - N_{\epsilon_{n,l}}(N-1)   \ne 0 .
\end{displaymath}
In practice, we do not consider all individual $\Delta N_{\epsilon_{n,l}}$, 
but only the sums 
\begin{displaymath}
  \Delta N_l = \sum_n \Delta N_{\epsilon_{n,l}} .
\end{displaymath} 

For a given angular momentum $l$, the first level to be filled as the cluster 
size increases will be the one with the nodeless radial function: 
$\epsilon_{0,l}$, followed by those with quantum numbers $n=1,2,\ldots$. 
Since these levels are well separated, their filling
will be indicated by small humps in $\Delta N_l$, which otherwise vanishes. 
Fig.~\ref{occs} shows the $\Delta N_l$, shifted vertically by $l$, as 
functions of $N^{1/3}$.  

Since orbitals having the same energy are filled simultaneously, finding 
several humps for the same cluster size means that the corresponding orbitals
$\epsilon_{n,l}$ are degenerate. For the the leading levels (i.e.\ those with 
large angular momentum) these degeneracies are indicated in Fig.~\ref{occs}.
The patterns are similar for all jellium densities. 

\noindent
\begin{minipage}{3.375in}
\begin{figure}
  \centerline{\epsfxsize=3.37in \epsffile{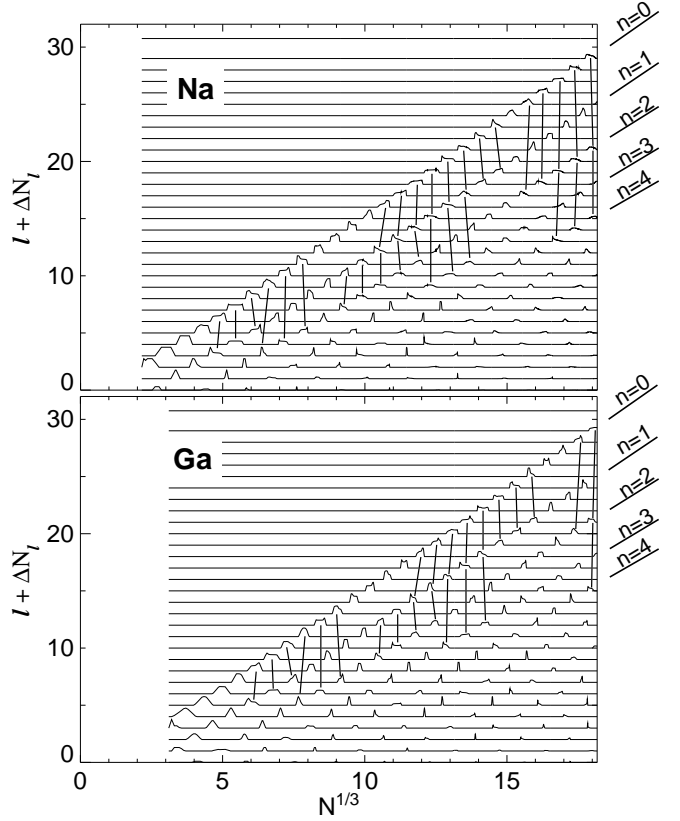}}
  \vspace{1ex}
  \caption[]{\label{occs}
             Filling of the orbitals with increasing cluster size $N$ in the 
             self-consistent calculations for jellium clusters. $\Delta N_l$ 
             is the change in the occupation of the orbitals with angular 
             momentum $l$. In the plot, the $\Delta N_l$ are shifted by $l$.
             The small humps in the horizontal lines indicate a change in the 
             occupation of an orbital with the corresponding angular momentum, 
             i.e., the filling of that orbital. Since degenerate orbitals are
             filled simultaneously, the existence of several such humps for a 
             given cluster size $N$ shows that the corresponding orbitals are
             degenerate. For the orbitals with large angular momentum these 
             degeneracies are indicated by lines connecting such humps.
             The first hump for a given $l$ corresponds to the orbital with 
             nodeless radial wave function $n=0$. Likewise, the orbital 
             connected with the second hump has the quantum number $n=1$. 
             As can be seen from the plots, the humps for orbitals with given 
             $n$ are arranged into nearly straight lines of slightly different 
             slope.}
\end{figure}
\end{minipage}

\vspace{2ex}
\noindent
The main difference is that the pattern is shifted towards 
larger cluster sizes as the
Wigner-Seitz radius decreases. We find that the first maximum in the shell
amplitude is caused by the degeneracy of the states $\epsilon_{0,l} \approx 
\epsilon_{1,l-3}$ followed by $\epsilon_{0,l} \approx \epsilon_{2,l-6}$. 
The second maximum in the supershell structure is mainly due to the degeneracies
$\epsilon_{1,l} \approx \epsilon_{2,l-3}$, supported by $\epsilon_{0,l} \approx 
\epsilon_{1,l-4}$ and $\epsilon_{1,l} \approx \epsilon_{3,l-6}$.

To find out why there are nodal regions in $\tilde{E}$, we first observe that 
the humps with a given quantum number $n$ are arranged into nearly straight 
lines (see Fig.~\ref{occs}). The slope of these {\em occupation lines} is 
larger for smaller $n$, i.e., the distance $\Delta l$ between two such lines 
increases with cluster size $N$. But since the angular momentum is 

\noindent
\begin{minipage}{3.375in}
\begin{figure}
  \centerline{\epsfxsize=3.37in \epsffile{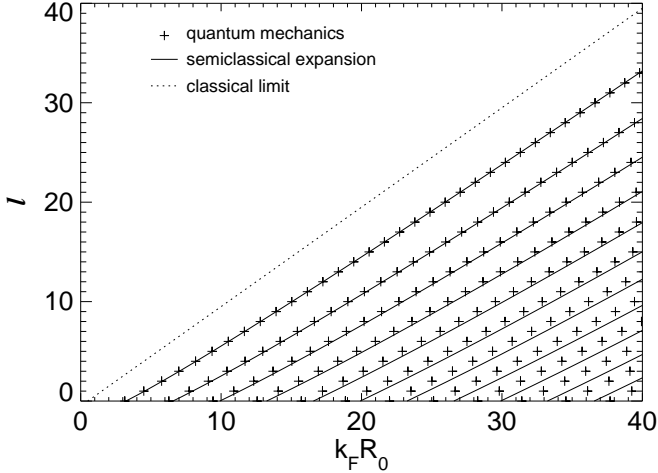}}
  \vspace{1ex}
  \caption[]{\label{occlines}
             Position of the energy levels $\epsilon_{n,l}=\hbar^2 k_{n,l}^2/2m$
             in an infinite spherical potential well (spherical cavity) of 
             radius $R_0$. For comparison with Fig.~\ref{occs}, we give 
             $k R_0$ on the abscissa and the angular momentum $l$ on the 
             ordinate. The position of the energy levels is indicated by 
             crosses. For small $n$, these can be well described by the 
             semiclassical expansion that is described in the text. The 
             classical limit $L_{max}=\hbar k_F R_0$ is given by the dotted 
             line.}
\end{figure}
\end{minipage}

\vspace{2ex}
\noindent
quantized, degenerate energy levels exist only for integer 
values of $\Delta l$. Therefore
regions where energy levels are degenerate are separated by regions without 
degeneracies. Hence the supershell structure can be traced back to the 
differences in the slopes of the occupation lines.

The origin of these lines can be understood from a simple classical argument.
Assuming that the Fermi energy $E_F$ is fairly independent of the cluster size,
the maximum angular momentum for an electron in a potential well of radius
$R_0=r_s\,N^{1/3}$ is given by
\begin{equation}\label{Lclass}
  L_{max}(N) = \hbar k_F\,R_0 ,
\end{equation} 
where $\hbar k_F=(2mE_F)^{1/2}$ is the Fermi momentum. Using semiclassical
quantization, we can go beyond this crude approximation. The angular momentum
is then written as $L=\hbar(l+1/2)$, where $l$ is the angular momentum quantum
number. Introducing the parameter
\begin{equation}
  \gamma = {l+1/2 \over k_F R_0} ,
\end{equation}
which is unity in the classical case (\ref{Lclass}), the Bohr-Sommerfeld 
condition for an infinitely deep potential well of radius $R_0$ reads
\begin{equation}\label{BScond}
 \Big(l+{1\over2}\Big)\left(\sqrt{{1\over\gamma^2} - 1} - \arccos(\gamma)\right)
 = \pi\Big(n + 3/4\Big) .
\end{equation}
Expanding this equation to first order around the classical case $\gamma=1$,
we can solve for the angular momentum of a state with quantum number $n$ at 
the Fermi level:
\begin{equation}\label{LBS}
  l+{1\over2} \approx k_F R_0
              - {\Big(3\pi(n+3/4)\Big)^{2/3} \over 2}\,(k_F R_0)^{1/3} .
\end{equation}
Fig.~\ref{occlines} shows the angular momentum quantum number $l$ as a 
function of $k_F R_0$. It turns out that, for small quantum numbers $n$, 
the approximation (\ref{LBS}) matches the quantum mechanical result quite well.
The deviations for larger $n$ are due to the expansion around $\gamma=1$, not
to the semiclassical approximation itself.
Since a state $\epsilon_{n,l}$ at the Fermi level will be filled as the cluster
size increases, the crosses in Fig.~\ref{occlines} can be identified with the
humps in Fig.~\ref{occs}. Hence the expansion (\ref{LBS}) provides, for small
$n$, a good approximation to the occupation lines for the infinite potential 
well.

\section{Scaling Analysis}\label{scaling}

In the present section we analyze the systematic variations
in the shell and supershell structure as the jellium density is changed. 
The key to such an analysis is the identification of relevant scales.
Expressing all data in terms of these scales, i.e.\ dealing with dimensionless
{\em reduced quantities}, allows the results for different materials to be
conveniently compared.

Obviously, the Wigner-Seitz radius $r_s$ is such a scale. It is a fundamental
length describing how the volume of the cluster depends on the jellium density.
Via the relation
\begin{displaymath}
  k_F \approx \sqrt[3]{9\pi\over4}\;{1\over r_s}
\end{displaymath}
it is also related to a basic energy scale. As a simple example we note that
the overall amplitude of the shell oscillations $\tilde{E}$ is proportional
to $1/r_s^2$. This is shown in Fig.~\ref{Eampl}. 

From the shape of the self-consistent Kohn-Sham potentials we can 
identify a second length scale. As can be seen from Fig.~\ref{shiftpot},
for a given jellium density the shape
\noindent
\begin{minipage}{3.375in}
\begin{figure}
  \centerline{\epsfxsize=3.37in \epsffile{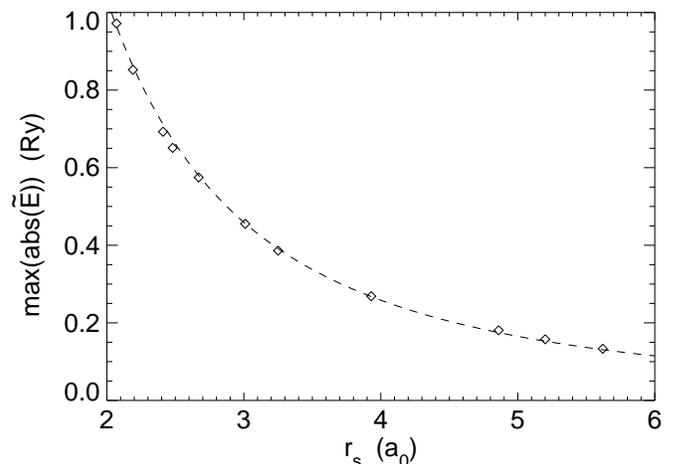}}
  \vspace{1ex}
  \caption[]{\label{Eampl}
             Scaling of the amplitude of the shell oscillations with the
             electron density $n_{el}=3/(4\pi r_s^3)$. The diamonds give the
             maximum amplitude of the oscillating part $\tilde{E}$ extracted
             from our self-consistent calculations. The dotted line gives 
             a fit of these data points with the one-parameter function
             $const./r_s^2$.} 
\end{figure}
\end{minipage}

\noindent
\begin{minipage}{3.375in}
\begin{figure}
  \centerline{\epsfxsize=3.37in \epsffile{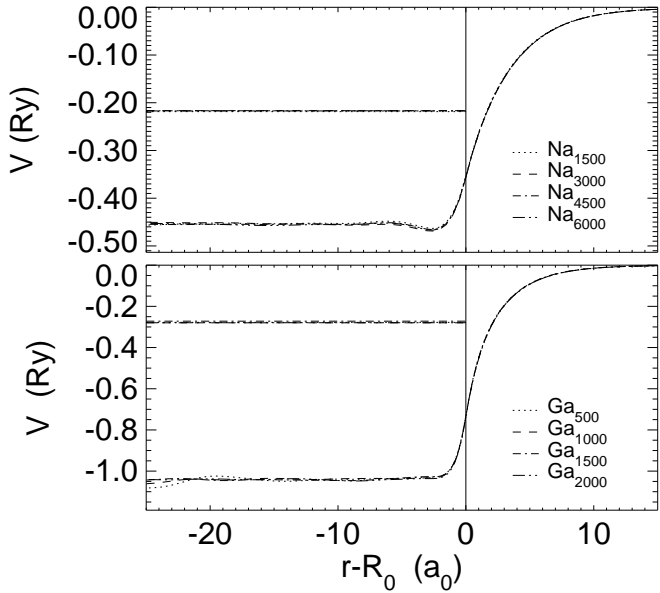}}
  \vspace{1ex}
  \caption[]{\label{shiftpot}
             Self consistent Kohn-Sham potential for sodium and 
             gallium jellies of different sizes. To compare the surface region
             of the potentials, we chose $r$ minus the radius $R_0$ of the
             corresponding jellium as the abscissa. As can be seen, the
             shape of the potential at the surface of the clusters is fairly
             independent of the size $N$. The horizontal lines indicate the
             Fermi energies of the corresponding systems.}
\end{figure}
\end{minipage}

\vspace{2ex}
\noindent
of the potentials around $R_0$ is quite 
independent of cluster size. For each material it can be characterized by the
width of the surface region, which hence is a good candidate for an additional 
length scale. We will denote the surface width by $a$.

We might assume then that the changes in the electronic shells and supershells
can be described in terms of the dimensionless parameter $a/r_s$. For a rough 
estimate of how $a$
varies with the jellium density, we have performed a 
simple Thomas-Fermi calculation for a plane jellium surface 
$n_I(z)=n_I\,\Theta(-z)$. Given an area $A$ the surface energy is
\begin{displaymath}
  \sigma[n_{el}(z)] 
   = {\int\limits_{-\infty}^\infty \Big(\epsilon(n_{el}(z))\,n_{el}(z)
                         - \epsilon(n_I(z))\,n_I(z)\Big)\,dz \over A} ,
\end{displaymath}
where $\epsilon(n)$ is the Thomas-Fermi energy functional including 
the second order gradient correction to kinetic energy,\cite{Kirzhnits} 
exchange energy and Hartree energy. To estimate $a$, we minimize the surface 
energy using a parametrized electron density, where the only parameter is the 
surface width
\begin{displaymath}
  n_{el}(z) = n_I\,f(z/a) .
\end{displaymath}
A good choice for the function $f(z)$ is\cite{smith69}
\begin{displaymath}
  f(z) = \left\{\begin{array}{cl}
                   1-{1\over2}e^z\quad &,\;z<0 \\
                     {1\over2}e^{-z}   &,\;z\ge0  \;.
                \end{array}\right. 
\end{displaymath}

\noindent
\begin{minipage}{3.375in}
\begin{figure}
  \centerline{\epsfxsize=3.37in \epsffile{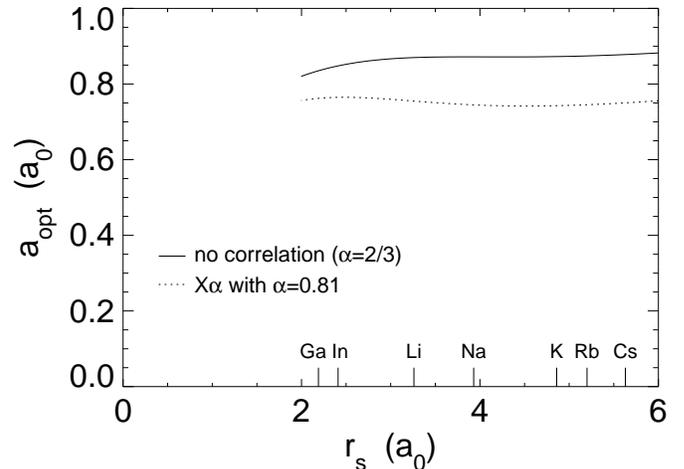}}
  \vspace{1ex}
  \caption[]{\label{surfTF}
             Width $a_{opt}$ of the surface region as a function of electron 
             density. $a_{opt}$ was obtained from a Thomas-Fermi calculations
             (including gradient corrections and exchange) for a parametrized 
             electron density. The energy functional included second-order
             gradient corrections to the kinetic energy, Coulomb energy, and
             exchange energy (full line). Including also exchange in the 
             $X_{\alpha}$ approximation (Ref.\cite{Slater4}) yields the dotted 
             curve.}
\end{figure}
\end{minipage}

\vspace{2ex}
The resulting surface width $a_{opt}$ as a function of the Wigner-Seitz radius
$r_s$ is shown in Fig.~\ref{surfTF}. It is fairly independent of the jellium 
density, in agreement with the results of the self-consistent Kohn-Sham 
calculations shown in Fig.~\ref{shiftpot}. Hence $r_s$ seems to be the only 
relevant length scale in the cluster problem.

The dependence of the shell and supershell structure on the scaling parameter
$a/r_s$ is revealed by plotting the magic numbers and the supernodes against 
$1/r_s$. Fig.~\ref{mag_plot} shows the resulting pattern, which is 
surprisingly simple. The magic numbers are practically independent of the 
jellium density, while the supernodes are linearly shifted with $1/r_s$. 
A closer look at the supernodes reveals that their position can be accurately
interpolated by a linear function. For some given jellium density a reliable 
estimate of the location of the minima in the supershell structure can be
read off Fig.~\ref{jellynod}, without having to perform a time-consuming 
self-consistent calculation.

It is interesting to note that the first and the second supernode
are shifted by the same amount. Writing the oscillating part of the total 
energy as a beating pattern 
\begin{eqnarray}
  \tilde{E}(N)
&=& \cos\Big(f_1 N^{1/3}-\varphi_1\Big) + \cos\Big(f_2 N^{1/3}-\varphi_2\Big)
    \nonumber \\[1ex]
&=& 2\;\cos\Big({f_1-f_2\over2} N^{1/3} - {\varphi_1-\varphi_2\over2}\Big)
    \label{SuperOsc} \\
&&\times\cos\Big({f_1+f_2\over2} N^{1/3} - {\varphi_1+\varphi_2\over2}\Big) 
    \nonumber 
\end{eqnarray}
suggests a simple picture for understanding such a parallel displacement. 
Since the supershell oscillation is described by the first cosine in 
(\ref{SuperOsc}), the position of the $k^{th}$ supernode is given by

\noindent
\begin{minipage}{3.375in} 
\begin{figure}
  \centerline{\epsfxsize=3.37in \epsffile{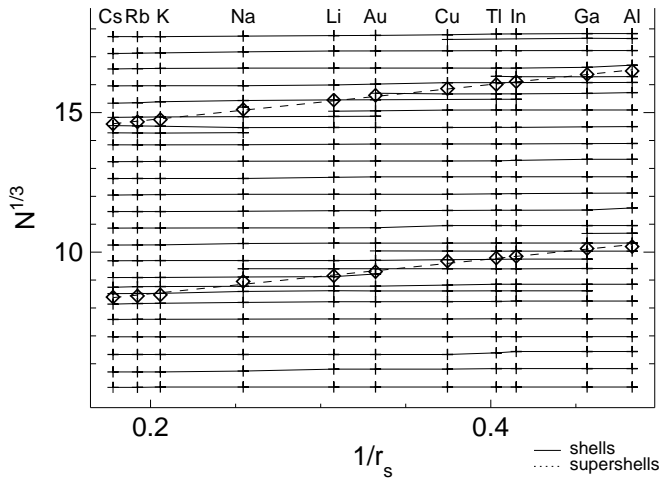}}
  \vspace{1ex}
  \caption[]{\label{mag_plot}
             Position of the shell minima (crosses) and of the supernodes
             (diamonds) as a function of $1/r_s$. Corresponding shell minima
             and supernodes are connected by full and broken lines, 
             respectively.} 
\end{figure}
\begin{figure}
  \centerline{\epsfxsize=3.37in \epsffile{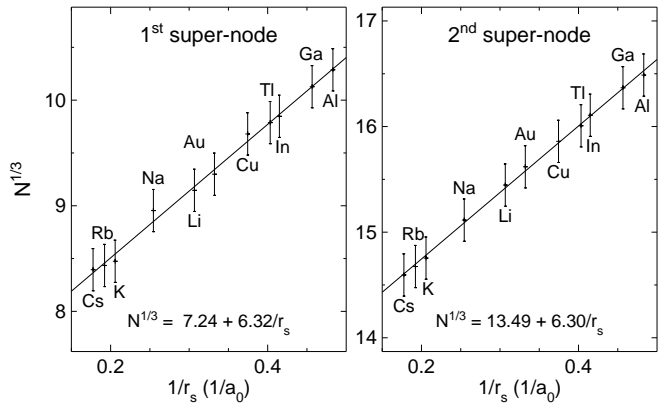}}
  \vspace{1ex}
  \caption[]{\label{jellynod}
             Shift of the first and second supernode as a function of $1/r_s$.
             The error bars are due to the uncertainty in determining the 
             the envelope of $\tilde{E}(N)$. The lines give a linear fit of 
             the data, the parameters of which are displayed in the plot.} 
\end{figure}
\end{minipage}

\begin{displaymath}
  N^{1/3}_k = { (2k+1)\pi\over f_1-f_2} + {\varphi_1-\varphi_2\over f_1-f_2} .
\end{displaymath}
In this picture a shift of the supernodes, which leaves the period of the
oscillation unchanged, naturally arises from a variation of the phases
$\varphi_1$ and $\varphi_2$. In a semiclassical approach, using a periodic
orbit expansion,\cite{nishioka90} these phases are determined by the
cluster surface. This is consistent with the observation that the supershell 
structure is sensitive to the shape of the surface 
potential.\cite{clemenger91,Lerme93b} 
Using an expansion of the relevant semiclassical integrals in terms of the
scaling parameter $a/r_s$ (leptodermous expansion) it can be shown that the
above interpretation indeed describes the mechanism underlying the observed
shift of the supernodes.\cite{dissl}

\section{Conclusions}

We have analyzed how the oscillating part $\tilde{E}$ of the total energy
extracted from self-consistent calculations depends on the electron density.
It turns out that the magic numbers are fairly independent of the density,
while the supershell oscillations are shifted towards larger cluster sizes
as the Wigner-Seitz radius decreases. Focusing on the energy spectrum near
the Fermi level $E_F$, we have found a simple pattern in which the orbitals 
are filled: the occupation lines. They can be understood using a semiclassical 
expansion. Qualitatively there is no difference between clusters of different 
density. The strong oscillations in $\tilde{E}$ are caused by degeneracies of 
energy levels close to $E_F$. The supershells are a consequence of the 
different slopes of the occupation lines, which imply transition zones 
between regions where levels are degenerate.
The shift of the supernodes with increasing density can be understood from
an analysis of the physical scales. We have identified $r_s$ as the relevant 
length scale for the cluster problem. It turns out that the location of the
supernodes is linearly shifted with $1/r_s$.

Furthermore, the identification of $r_s$ as the typical length scale for
the supershells suggests a justification of the {\it ad hoc} procedure
proposed in Ref.\cite{Broyer93} to improve the results of jellium calculations
for gallium clusters. There it was found that the introduction of a
non-homogeneous jellium background is essential for treating Ga$_N$
clusters, while alkali-metal clusters are well described by a homogeneous
jellium. With the help of the above scaling argument that puzzle may be 
resolved. Assuming that the typical length scale for features in the
jellium is the ionic radius $r_{at}$, while the length scale for the electrons
is the Wigner-Seitz radius $r_s$, we find that the importance of
inhomogeneities increases with the number of valence electrons 
$Z_{val} \propto (r_{at}/r_s)^3$. Hence, for trivalent materials a soft 
jellium surface should have a stronger influence on the supershells than 
for alkali metals.

\section*{Acknowledgments}

Helpful discussions with T.~P.~Martin and M.~Brack are gratefully acknowledged.
M.~R.~Pederson, D.~D.~Johnson, J.~P.~Perdew, and P.~Dufek provided us with 
subroutines for Broyden-mixing\cite{Broyden} and gradient corrections. 
Without the continued support of A.~Burkhardt the extensive jellium 
calculations would not have been possible.

\bibliographystyle{prsty_long}

\end{multicols}
\end{document}